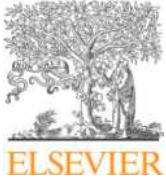
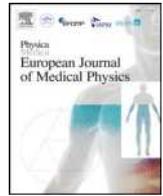

Original paper

# Shaping of a laser-accelerated proton beam for radiobiology applications via genetic algorithm

M. Cavallone[a], A. Flacco[a,*], V. Malka[a,b]

[a] *Laboratoire d'Optique Appliquée, ENSTA-ParisTech, École Polytechnique, CNRS-UMR7639, Institut Polytechnique de Paris, 828 bd des Maréchaux, 91762 Palaiseau cedex, France*
[b] *Department of Physics of Complex Systems, Weizmann Institute of Science, Rehovot 7610001, Israel*

ABSTRACT

Laser-accelerated protons have a great potential for innovative experiments in radiation biology due to the sub-picosecond pulse duration and high dose rate achievable. However, the broad angular divergence makes them not optimal for applications with stringent requirements on dose homogeneity and total flux at the irradiated target. The strategy otherwise adopted to increase the homogeneity is to increase the distance between the source and the irradiation plane or to spread the beam with flat scattering systems or through the transport system itself. Such methods considerably reduce the proton flux and are not optimal for laser-accelerated protons. In this paper we demonstrate the use of a Genetic Algorithm (GA) to design an optimal non-flat scattering system to shape the beam and efficiently flatten the transversal dose distribution at the irradiated target. The system is placed in the magnetic transport system to take advantage of the presence of chromatic focusing elements to further mix the proton trajectories. The effect of a flat scattering system placed after the transport system is also presented for comparison. The general structure of the GA and its application to the shaping of a laser-accelerated proton beam are presented, as well as its application to the optimisation of dose distribution in a water target in air.

## 1. Introduction

The very short duration at the source (~ps) and the extremely high attainable peak dose rate (> $10^9$ Gy/s delivered during ~1 ns pulses) of laser-accelerated proton beams open new axes of research in radiation biology, aiming at investigating the biological response of living cells to fast dose deposition [1–5]. Laser proton acceleration is obtained by focusing a high-power laser (TW-PW) on a thin solid foil, thus creating a plasma. Several acceleration mechanisms can take place according to the different interaction regimes between the laser and the plasma [6]. In Target Normal Sheath Acceleration (TNSA) [7] high-energy electrons are produced at the illuminated surface and propagate through the target bulk, driving the plasma expansion. The high electric field associated (TV/m) accelerates protons coming from hydrogen-rich impurities in a cone along the axis perpendicular to the foil surface. Such proton beams feature an exponential energy spectrum with cut-off energies reaching several tens of MeV, and a large divergence around the propagation axis.

TNSA-driven proton beams are not suitable for radiobiology experiments without a proper transport system allowing spectral and spatial shaping. Different strategies using electro-magnetic devices have been studied and employed for *in vitro* irradiation experiments. Such solutions include the use of dipoles [8,9], energy selection systems (ESS) [10,11] permanent magnet quadrupoles (PMQs) for beam shaping [12] and combinations of them [13,14]. Also, more complex systems consisting of a compact proton gantry composed of ESS, PMQs and a pulsed solenoid have been recently designed and investigated to bring the laser-acceleration technology to the clinics in the future [15]. The feasibility of using such system to deliver a homogeneous dose over a 3D volume through an active scanning dose delivery approach has been also demonstrated [16]. In addition to the development of advanced dose delivery techniques, a parallel research focused on assessing the radiobiological effectiveness and consequences of fast dose deposition on living cells by laser-driven protons is necessary. Given the limited charge available in a laser-accelerated proton bunch and the inevitable losses in transport, one of the challenges in its application to radiobiology is to concentrate as many protons as possible on the biological sample, maintaining an acceptable level of dose homogeneity. In

M. Cavallone, et al.                                                                                                                                 Physica Medica 67 (2019) 123–131

Ref. [17] the use of flat scattering foils placed after the transport system has been proposed to deliver a transversal homogeneous dose over a sample with dimensions of clinical relevance (~cm$^2$) for future radiobiology studies at the ELIMED facility [18,19]. In other experiments, a sufficient transversal homogeneity was obtained by spreading the beam with the transport system itself.

In this paper we propose a method to shape a laser-driven proton beam through the use of a scattering system (SS) composed by tiles of variable thicknesses. We show that placing such scattering system before a quadrupole element allows a more efficient shaping of the beam, improving the trade-off between total flux and transversal beam homogeneity at the sample. The design of such scattering system is optmised by a Genetic Algorithm (GA). GAs are heuristic computing techniques that have been developed to find the best or near-best solution to optimization problems where the parameters are numerous and a theoretical approach is not suitable. Their application area is wide and includes engineering [20], traffic problems [21], medical applications such as radiology, radiotherapy, oncology and surgery [22]. This type of algorithm makes use of an abstract version of the Darwinian laws of genetics: a population of possible solutions, which are called chromosomes, is recombined randomly and evolves from a generation to the next through an iterative process. A selection of the fittest solutions is operated at every iteration with the use of a fitness function that contains the parameters to be optimized. The selected solutions are recombined randomly to give rise to a new generation of fittest solutions, which go again through a selection and recombination process. The loop is repeated until a certain stop criteria is met, for example when a 'good-enough' solution is found or when a maximum number of iteration is reached. A detailed description of GAs, which is beyond the scope of this paper, can be found in McCall (2015) [23].

The optimisation of the scattering system design involves two steps. In a first step, Monte Carlo (MC) Geant4-based simulations are performed to simulate the beam transport, the interaction with matter, and to retrieve the beam parameters at the scattering system and at the irradiated sample. These simulations are needed to produce a base of elementary dose and flux maps that are used as input to the following optimisation problem. Secondly, a GA is employed to optimise the design of the scattering system according to given requirements.

## 2. Optimisation method

A schematic drawing of the design process is shown in Fig. 1. The first step involves MC simulations to reproduce the propagation of the beam through the transport system and in particular the interaction with a scattering system. The scattering system is a grid of tiles of variable discrete thicknesses, whose effect on the beam is simulated through the MC simulations. Before starting the design process, the values and number of discrete thicknesses that will be used for the design need to be determined. The choice depends on many factors, such as the energy spectrum of the beam and the computational burden. In fact, a number of MC simulations equal to the number of available thicknesses needs to be performed. The aim is to correlate the effect of every tile, for each possible thickness placed in every position of the scattering system grid, on the 2D flux and dose profiles at the target. Therefore, in each MC simulation a scattering foil of a different thickness and a target are placed at a well-defined position. During the simulation, the propagation of each primary particle, crossing the scattering system at the tile (i,j) and arriving to the target plane at the quantized coordinates (k,l), is recorded. The dose deposited in the target is then added to the cell (i,j,k,l) of a 4-dimensional matrix. In other words, elementary flux and dose maps at the target are created separately for each tile of the SS crossed by a different portion of the beam, as shown in the schematic drawing of Fig. 1. Once all the MC simulations are performed, the output at the target can be reconstructed for any scattering system design made up of different discrete values of thickness in each tile.

In the second step, a GA is used to find the optimal configuration of the scattering system and meet the required specifications at the target. The GA is initialised with the creation of the first generation of $N = 500$ random scattering systems. The resulting maps at the target associated to each design are reconstructed using the data set of elementary maps previously obtained with MC simulations. Such 2D profiles are then evaluated through the use of a fitness function, which associates a scalar value to each of the analysed solutions, depending on how good a scattering system performs with respect to the pre-defined criteria for the desired output. After the evaluation phase, the scattering systems under analysis undergo the recombination phase. In the recombination phase, N-1 parent couples of scattering systems are randomly selected among the N/2 fittest ones. From each parent couple a new child scattering system is created by randomly mixing the parent tiles' thicknesses. At the end of this phase, a new population of scattering systems is composed by the N-1 issued by recombination plus the fittest one promoted from the previous population. This ensures the fitness function to remain monotonic non-decreasing over GA iterations. The new population is then fed to the GA algorithm for the next iteration of the loop, which is repeated until the algorithm converges to a solution.

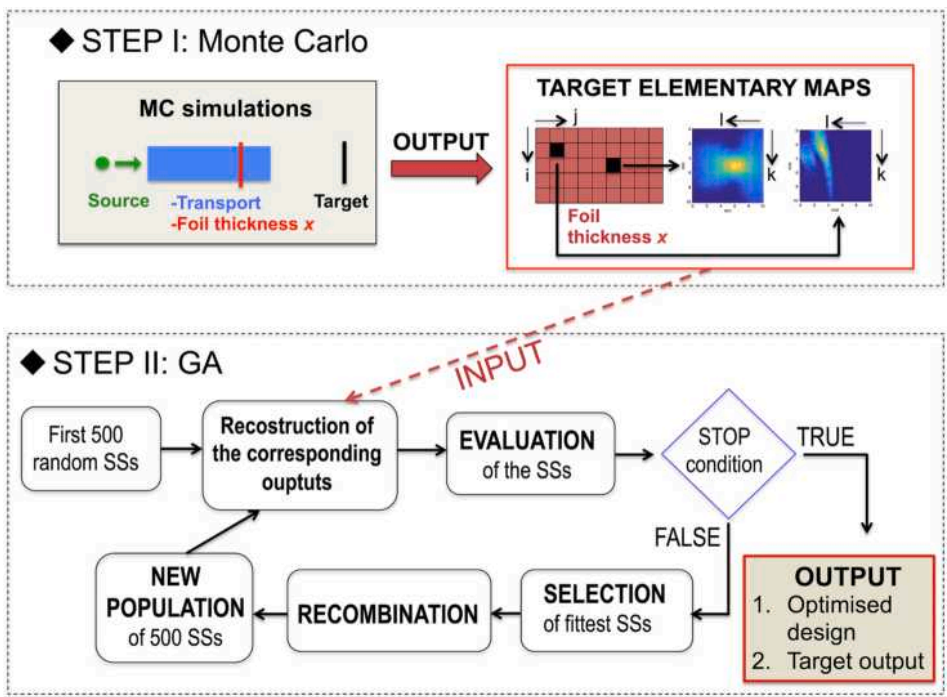

**Fig. 1.** Scheme of the scattering system design process. Monte Carlo simulations (STEP I) are first performed to produce flux and dose elementary maps that are used as input in the Genetic Algorithm (GA) (STEP II) to optimise the scattering system (SS) design. The elementary maps at the target are retrieved separately for each portion of the beam crossing a different tile of the SS. The final outputs obtained with the GA are the optimised SS design and the corresponding optimised output maps at the target.

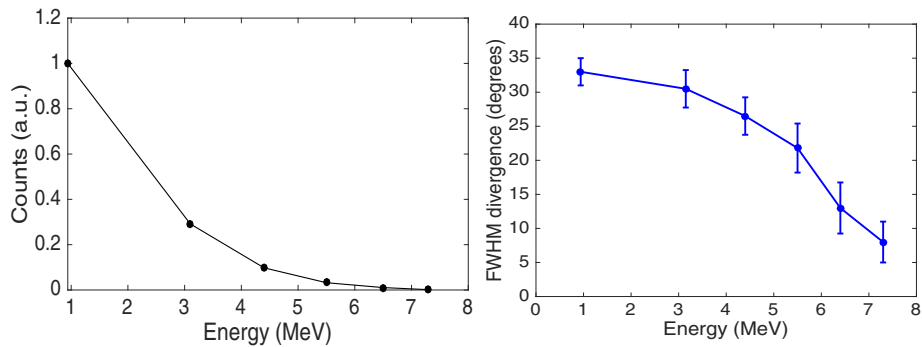

**Fig. 2.** Energy spectrum and divergence of the proton beam at the source (the plot on the right is taken from Pommarel et al. (2017) [12]).

### 3. Application to a laser-accelerated proton beam

This strategy is now applied to shape the laser-accelerated proton beam used for *in vitro* irradiation at the LOA [5,12]. The proton source has a TNSA-like energy spectrum with a cut-off at 7.5 MeV and a Gaussian angular divergence around the propagation axis that is a function of the energy, as shown in Fig. 2. The spectrum and the divergence were obtained using a radiochromic film stack placed 5 cm after the source (more details are given in Ref. [12]). The beam is guided towards a Mylar window by a set of four quadrupoles validated during previous campaigns [24] to focus the spectral bandwidth around 5–6 MeV. The envelopes of different energetic components of the beam along the transport system as well as the final spectrum focused after the transport system are shown in Fig. 3. The beam transport and the interaction with matter are simulated using the G4beamline package (release 3.04) [25], a Monte Carlo particle-tracking program based on Geant4 (Geant4.10.04 version) [26,27] specifically developed to simulate beamlines and transport systems. With respect to Geant4, additional collective computations such as space charge effect are implemented. The quadrupoles were simulated using measured field maps [28] and the proton source was reproduced using the energy spectrum and the energy dependent angular divergence shown in Fig. 2. The physics list employed is the QGSP_BIC, which has been extensively validated in the energy domain of medical applications for proton beams [29,30]. As a single particle MC code, Geant4 cannot natively simulate space charge effect, which motivated the choice of using a modified package like G4beamline. As far as the configuration of our transport system is concerned, it turned out that space charge effects remain negligible. The largest part of simulations for this manuscript was performed on a modified version of G4beamline (to make the output data format compatible with the GA code) without space charge effects. The beam propagation simulated using G4beamline showed perfect agreement with results obtained using the Hadrontherapy Geant4-based tool [31,32]. Also, the beam propagation simulated with both tools proved to be consistent with EBT3 recorded profiles [24].

#### 3.1. Beam shape optimisation in vacuum

The GA is used to design a scattering system to shape the beam flux profile over a $1\,cm^2$ target surface placed 10 cm after the last quadrupole in vacuum, so as to avoid any scattering effect in air and on the Mylar exit window. The profile of the beam at this position without any scattering system placed along the beamline is shown in Fig. 4. The peaked profile is due to the Gaussian angular divergence at the source. Three configurations have been analysed:

1. Scattering system placed before the third quadrupole
2. Scattering system placed before the fourth quadrupole
3. Scattering system placed after the fourth quadrupole

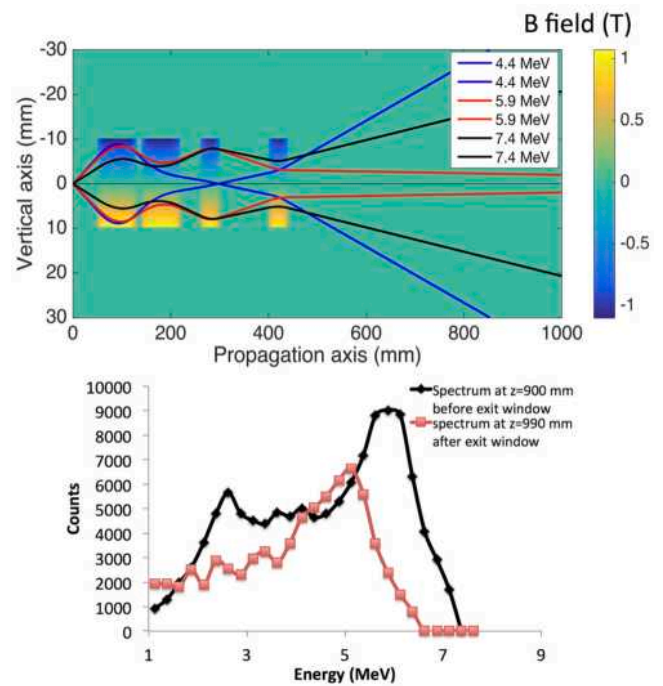

**Fig. 3.** Top: envelopes of different energetic components of the beam along the transport system. The two sets of two quadrupoles have a net bore aperture of 20 mm and a length of respectively 80 mm and 40 mm, with an average magnetic field gradient of about 100 T/m [24]. The trajectories are obtained using a matrix approach and an average value of the B gradient in the quadrupoles. Bottom: spectra over a $1 \times 1\,cm^2$ surface obtained with Geant4 simulations using $10^7$ particles. In black, spectrum at z = 900 mm. The charge reaching the surface is 1.2% of the charge at the source. In red, spectrum at z = 990 mm, 5 cm after a Mylar window placed at z = 940 mm.

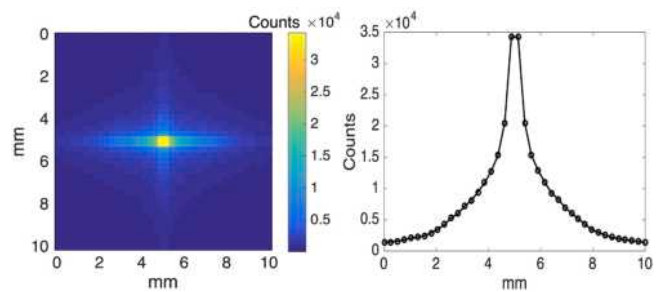

**Fig. 4.** Left: 2D flux profile on a surface placed 10 cm after the last quadrupole. Right: profile on the horizontal axis.

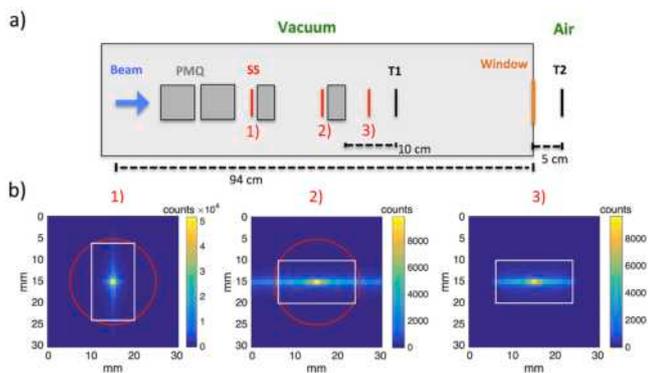

**Fig. 5.** a) Schematic drawing of the three configurations, b) beam shape at the scattering system position in the three configurations. The white rectangles represent the contour of the scattering system and the red circles represent the aperture of the quadrupole placed after it.

The results are also compared with the case of a simple flat scattering foil placed after the fourth quadrupole. The three configurations and the beam shapes at the scattering system positions are shown in Fig. 5. The scattering system is a $10 \times 18\,\text{mm}^2$ Mylar rectangle composed by 5x9 square tiles. The choice of Mylar was motivated by its low density, which allows to use tiles of manageable thicknesses. The dimensions and the orientation of the scattering system were chosen so as to cover the beam profile in the three configurations. The choice of the number of tiles was motivated by considerations on the number of particles that need to be simulated in MC simulations to retrieve input elementary maps for the GA with accuracy. For each geometry shown in Fig. 5, G4beamline simulations using $5 \times 10^7$ particles were performed to simulate the effect of 11 foil thicknesses going from 1 to 100 μm on the flux distribution at the target (T1) surface. The GA was then used to optimise the design of the scattering system to obtain the best trade-off between flux homogeneity and total flux at the target. The parameter employed to quantify the homogeneity is the standard deviation ($\sigma$) of the 2D profile. In this case, the fitness function that was minimised and used to select the scattering systems during the iterations was the following:

$$F = a \cdot \left( \frac{\sigma_{ss} - \sigma_{wss}}{\sigma_{wss}} \right) - b \cdot \left( \frac{\Phi_{ss} - \Phi_{wss}}{\Phi_{wss}} \right) \tag{1}$$

The first term in brackets represents the difference between the standard deviation of the flux at the target with and without scattering system and the second term in brackets represents the difference between the total flux at the target with and without scattering system. The two coefficients *a* and *b* are used to set different relative weights to the parameters. To compare the four configurations, the coefficients were adjusted in each case so as to obtain the same total flux at the target surface. Fig. 6 shows the optimised scattering system designs obtained with the GA for the three configurations and the corresponding flux at the target surface, as well as the flux obtained with a flat foil of 60 μm. For each case, the standard deviation is reported as a percentage of the mean value of the flux distribution.

As expected, the GA-designed scattering system placed after the transport system allows a better trade-off between the two parameters compared to a flat foil placed in the same position. This is due to the optimisation of the scattering effect over the transversal section of the beam obtained with the GA design: the beam is spread mainly at the centre and at the edges so as to flatten the profile in a more efficient way compared to a flat foil. A major improvement in homogeneity is obtained in the two first configurations, in which the scattering system is placed along the quadrupole system. In particular, the scattering system placed before the fourth quadrupole allows a decrease of 42% in standard deviation compared to the flat foil configuration. This is explained by the fact that such a scattering system allows the shaping of the beam in two ways: it spreads the beam through scattering effect and it intermingles the trajectories in the following quadrupole by introducing different energy losses on the transversal section of the beam and taking advantage of the quadrupole chromaticity. To give an example, Fig. 7 shows the effect of an energy loss of 0.5 MeV introduced before the last quadrupole on the external trajectories of the 5.5 MeV energetic component of the beam on the horizontal axis when neglecting the scattering effect. Without any energy loss the trajectories do not overlap, while by introducing an energy loss the trajectories are mixed and the profile is modified.

### 3.2. 2D dose optimisation for in vitro irradiation

This work was motivated by the need of new solutions to efficiently flatten the transversal dose distribution on in vitro biological samples and approach the constraints required in radiation biology [33,34], while keeping at the same time an acceptable value of dose per shot. The GA was therefore used to optimise the transversal dose distribution on a biological sample, represented by 10 μm thick water target with a $10 \times 10\,\text{mm}^2$ square surface placed 5 cm after the Mylar window in air (represented by T2 in Fig. 5). The percentage standard deviation is the parameter employed to quantify the flatness of the transversal dose distribution and compare the effect of GA-designed SS with flat foils. The dose distribution in the sample without any scattering system placed along the beamline is shown in Fig. 8, where the dose is expressed in Gy per 1 nC charge at the source, which was the estimated charge during experiments [12]. A maximum dose of 4 Gy/nC is delivered in the sample with a standard deviation of 77%, which is not suitable for radiobiology experiments. The GA was used to optimise the scattering system design in the 3 configurations already discussed. The fitness function used in this case was the following:

$$F = a \cdot \left( \frac{\sigma_{ss} - \sigma_{wss}}{\sigma_{wss}} \right) - b \cdot \left( \frac{D_{ss} - D_{wss}}{D_{wss}} \right) \tag{2}$$

The first term in brackets represents the difference between the standard deviation of the 2D dose distribution in the sample with and without scattering system and the second term in brackets represents the difference between the mean value of the 2D dose distribution in the sample with and without scattering system. Also in this case, we found that a non-flat scattering system placed before the fourth quadrupole flattens the transversal dose profile more efficiently than a flat foil placed downstream the beamline. In Fig. 9a we compare the standard deviation obtained using flat foils of increasing thicknesses placed after the fourth quadrupole with the standard deviation obtained using GA-designed scattering systems placed before the fourth quadrupole allowing the same maximum delivered dose. A 20–40% decrease in standard deviation is obtained with a genetic approach. Fig. 9b shows the scattering system design for the third case listed in the table and a comparison between the corresponding dose distribution and the dose distribution obtained with the 80 μm flat foil.

### 3.3. Sensitivity to shot-to-shot fluctuations

In view of experimental applications of this non-flat scattering system, shot-to-shot variations of the proton source need to be considered by studying to what extent they affect the beam shape and spectrum at the scattering system position and at the output target. We studied the effect of a 10% reduction in the cut-off energy (from 7.5 MeV to 6.75 MeV) as well as a change in beam pointing at the source of 5 mrad. Such values are based on observed instabilities during previous experimental campaigns at the SAPHIR facility. The effect of such instabilities on the dose delivered in a transmission ionisation chamber (TIC) employed during experiments was also quantified through Geant4 simulations. A variation of 25% in the delivered dose is caused by a 5 mrad change in pointing and a 10% reduction of the cut-off energy, which is consistent with the 15% typical standard deviation of the charge collected by the TIC during a day experimental run [12]. The study is performed by comparing the effect of such instabilities on the GA-designed scattering systems placed before the fourth quadrupole (designs shown in Figs. 6 and 9), which provide the best optimisation results in normal condition, and on the flat foils used for the comparisons. To simulate the spectrum instability, we rescaled the 7.5 MeV spectrum to a 6.75 MeV cut-off spectrum by decreasing both the minimum energy and the cut-off energy by a factor of $6.75/7.5 = 0.9$ and assuming the same available charge at the source (using a similar approach presented in Ref. [35,15]). The scaled spectrum is shown in Fig. 10 together with the original 7.5 MeV spectrum. In Fig. 10 we also show the spectra of the beam reaching the scattering system placed before the fourth quadrupole for the two cases. The reduction of the cut-off energy produces a loss of the portion of the spectrum above 6.75 MeV as well as a slight change in the relative yield of the energy components. The spectrum reaching the scattering system is peaked in a energy region that is lower than the cut-off energy, hence it is not affected much by such variation. However, when placing a GA-design scattering system in such position to optimise the output on a target surface, also the spectrum at the target needs to be considered. In fact, the final scattering system design is the result of the optimisation of scattering effect and energy losses on the portion of the spectrum that is focused on the target surface. The spectrum focused by the transport system at T2 (Fig. 5) is peaked in a region around 6 MeV (Fig. 3), therefore a decrease in the dose delivered with the 6.75 MeV spectrum at the source is expected both when using a GA-design scattering system and a flat foil. To assess the robustness of this approach, the key condition to verify is that the GA-designed scattering system is not less efficient than a flat scattering system placed after the transport system for such spectrum variations. In Table 1 are shown the results obtained at T1 and T2 for a 6.75 MeV cut-off spectrum at the source when placing before the fourth quadrupole the scattering systems designed for the 7.5 MeV case, and when placing after the transport system the flat foils used for the comparison (respectively 60 μm and 80 μm). As expected, a decrease in the dose reaching T2 is obtained both when using the GA-designed scattering system and the flat foil, compared to the 7.5 MeV cut-off case. This is due to the shift of the spectrum towards lower energies, which in turn causes a decrease in the number of protons reaching the target, whose spectrum is peaked around 6 MeV (Fig. 3), and an increase of the scattering effect. The change in the spectrum shape reduces the scattering system performance in terms of standard deviation at the target, yet the GA-designed scattering system still flattens the beam profile more efficiently than the flat foil. This is more evident for the flux optimisation at T1 in vacuum, in which the effects of a decrease in the cut-off energy are less important, the spectrum at this position being peaked at lower energies and because of the absence of other scattering elements such as the exit window and the air.

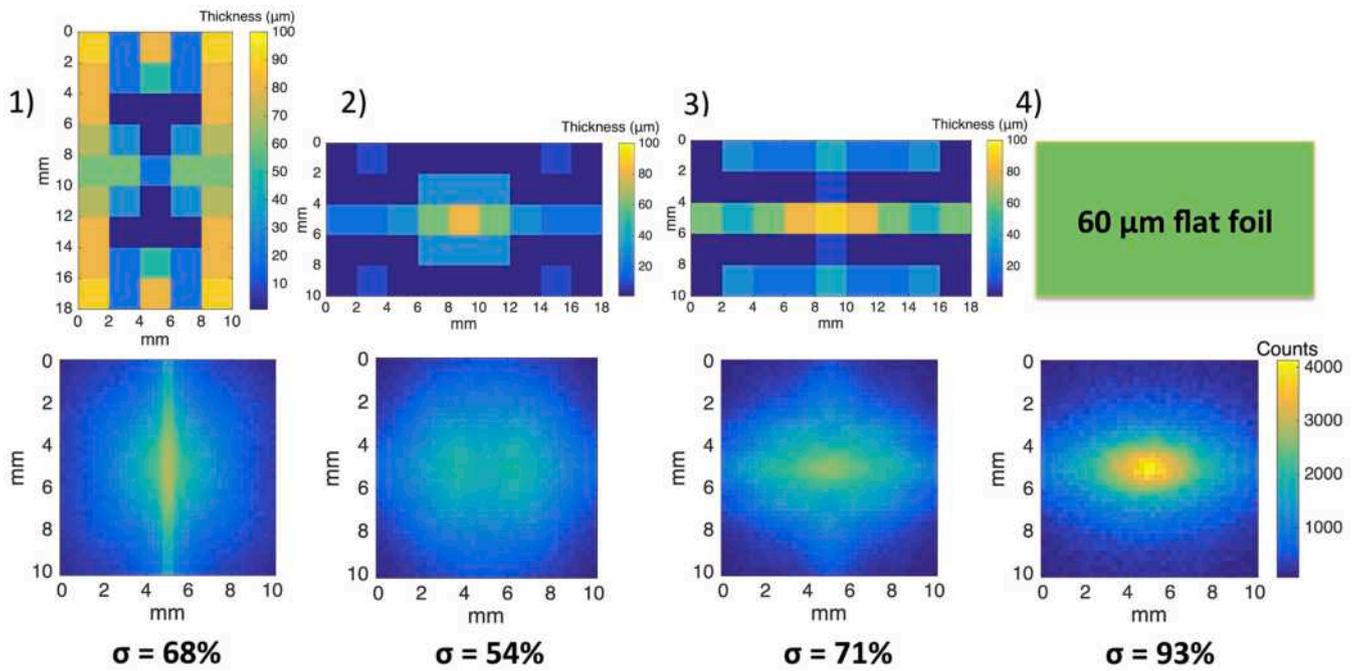

**Fig. 6.** Top: scattering system designs in the four configurations. Bottom: corresponding flux at the target surface (represented by T1 in Fig. 5). The total flux at the surface is the same for all configurations (2.9% of the protons at the source).

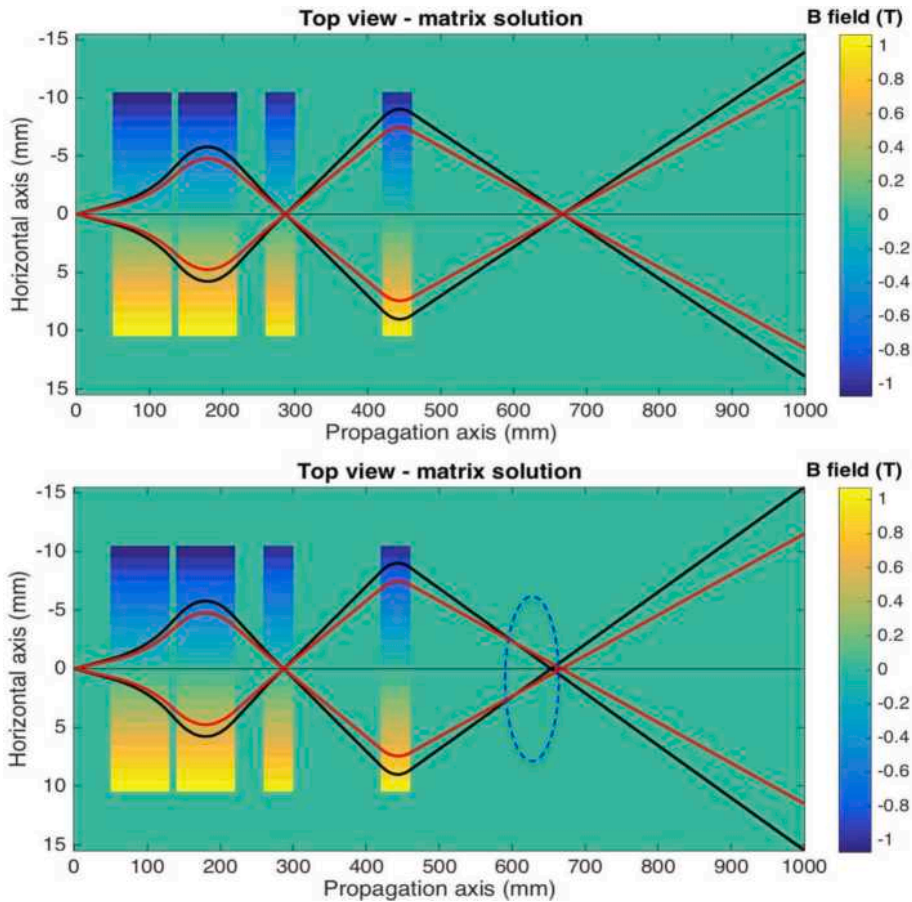

**Fig. 7.** Top: Trajectories of 5.5 MeV protons having two different values of the divergence at the source along the quadrupole system. The quadrupoles are simulated as ideal quadrupoles with an average value of the B gradient. Bottom: an energy loss of 5.5 MeV is introduced on the external trajectories (in black) before the fourth quadrupole entrance. In the region around z = 600 mm the trajectories are mixed.

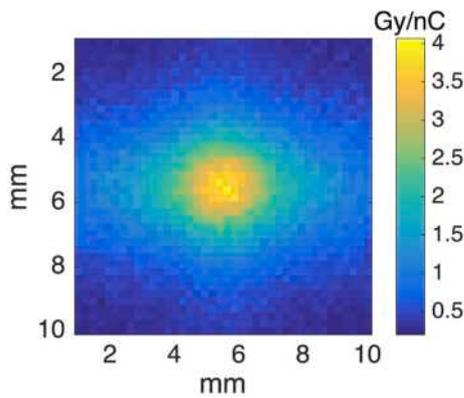

**Fig. 8.** Dose distribution over a surface placed 5 cm after the Mylar window without any scattering system along the beamline. The dose is expressed in Gy for 1nC charge at the source.

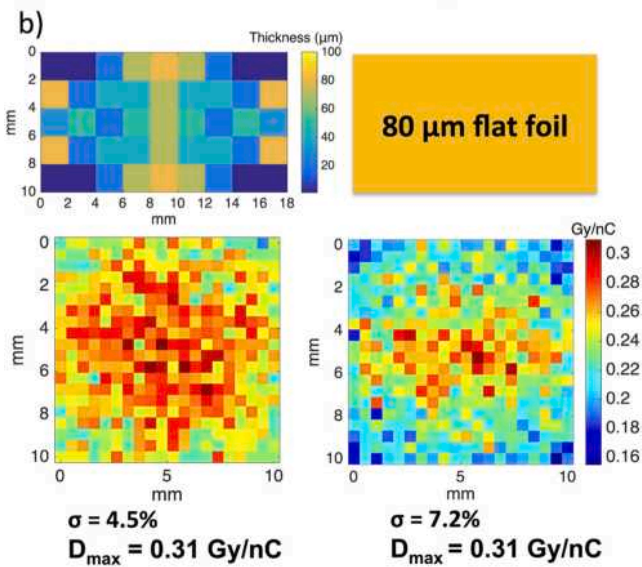

**Fig. 9.** a) Doses and standard deviations obtained at the target T2 with flat foils of increasing thickness compared with the standard deviation obtained using a GA-designed scattering system placed before the fourth quadrupole allowing the same delivered dose. b) Design of the non-flat scattering system and comparison between the dose distributions for the third case listed in the table.

Variations in beam pointing may also affect the dose distribution at the target and the scattering system efficiency. We studied the sensitivity to beam pointing in case of a 5 mrad change in the polar angle $\theta$ (in spherical coordinate system) and for an azimuthal angle $\phi = 45°$, which means the same deviation from the beam axis both on the horizontal and vertical plane. In Fig. 11a are shown the trajectories of some energetic components of the beam with the 5 mrad shift in pointing at the source. Such trajectories can be interpreted as the center of the corresponding energy envelope. At the T2 position ($z = 990$ mm) the transport system does not mitigate an initial misalignment on the horizontal plane for the 6 MeV component, which is representative for the spectrum reaching T2 (see Fig. 3), therefore an instability of the dose at the target is expected regardless the presence of a scattering system in the beamline. The beam shape before the fourth quadrupole with a 5 mrad shift of pointing at the source is shown in Fig. 11b (right). At this distance from the source ($z = 398$ mm), the change in beam position without any transport system would be of 1.4 mm on both axis while with the transport system the change is of about 2 mm on the horizontal axis and less than 50 μm on the vertical axis. Therefore, at this position the transport system mitigates the initial misalignment only on the vertical axis while it increases the misalignment on the horizontal axis, as it can be seen in Fig. 11. Moreover, 16% of the flux at this position is lost and the beam is blurred compared to the condition without misalignment (see Fig. 5). Considering that the scattering system is composed by square tiles of 2 mm size, such variation in beam position and shape is expected to affect the efficiency of a GA-designed scattering system placed in such position, as well as the dose distribution obtained with a flat scattering system placed after the transport system. In Fig. 12 are shown the dose distributions obtained at T2 when employing both the GA-designed scattering system and the flat scattering system. The distributions are also compared with the results obtained with no pointing misalignment, which were shown in Fig. 9.

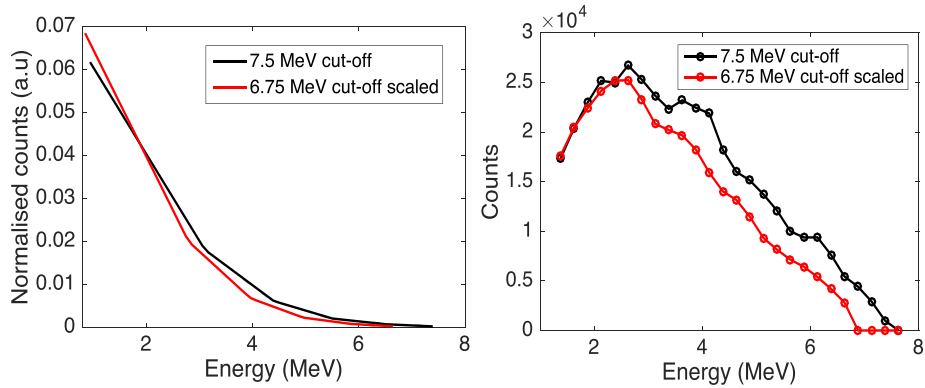

**Fig. 10.** Left: scaled spectrum at 6.75 MeV compared with the 7.5 MeV spectrum. Right: spectra at the scattering system placed before the fourth quadrupole for the two cases, obtained through Geant4 simulations using $10^7$ particles.

**Table 1**
Comparison between the GA-designed scattering system and the flat foils for the 6.75 MeV cut-off spectrum at the source. In brackets are reported the results previously obtained with the 7.5 MeV cut-off spectrum.

|  | GA-designed scattering system | Flat foil |
| --- | --- | --- |
| 2D flux at T1 | Flux = 2.6% (2.9%) <br> $\sigma$ = 61% (54%) | Flux = 2.5% (2.9%) <br> $\sigma$ = 92% (93%) |
| 2D dose at T2 (Gy/nC) | $D_{max}$ = 0.27 (0.31) <br> $\sigma$ = 5.8% (4.5%) | $D_{max}$ = 0.23 (0.31) <br> $\sigma$ = 6.8% (7.2%) |

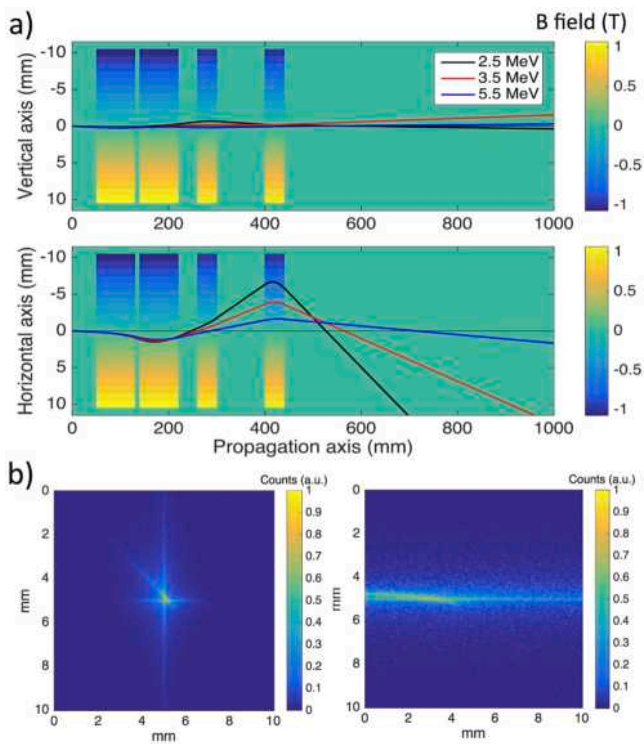

**Fig. 11.** a) Trajectories of protons with different energies with an initial misalignment of 3.5 mrad on both axis. The trajectories can be interpreted as the position of the center of the envelopes along the beamline. b) Beam shapes for a 5 mrad pointing misalignment. Left: beam shape before the third quadrupole. Right: beam shape before the fourth quadrupole. The shapes at the same positions with no misalignment were shown in Fig. 5.

When employing the 80 µm flat foil with an initial misalignment, the dose at the sample is slightly shifted compared to the situation with no misalignment, which causes a slight increase in the standard deviation. When employing the GA-designed scattering system, a slight increase in maximum dose and a major increase in standard deviation are obtained. This is due to the fact that the main part of the beam crosses the scattering system with an horizontal shift of ~2 mm, namely where a tile of a lower thickness is employed. Therefore more protons reach the target compared to the situation with no misalignment ($1.56 \times 10^7$ versus $1.4 \times 10^7$ for 1 nC charge at the source) because of the lowering scattering effect introduced. In conclusion, the dose distribution at the sample is affected in a different way in both cases. However, the sensitivity of the GA-designed scattering system can be reduced by placing it closer to a focus point, so as to find a better trade off between the beam spread at the scattering system surface, which needs to be large enough compared to the pixel size for the scattering system to be efficient, and the proximity to a beam focus point, which would allow a lower sensitivity to an initial misalignment. For example, the beam shape at the scattering system placed before the third quadrupole is less sensitive to an initial misalignment, as shown in Fig. 11b (left).

Also, the effect of a typical uncertainty in source position of 10 µm on the beam shape at the scattering system position was investigated. The results showed a misalignment at the scattering system of the same order of magnitude (less than 50 µm) and no appreciable change in beam shape.

## 4. Conclusions

This work reports on a proposed strategy to shape laser-accelerated proton beams that can be used in applications requiring high charge and high uniformity over a surface, such as radiobiology experiments. We demonstrated the use of a Genetic Algorithm assisted approach to design a non-flat scattering system optimally shaping the beam profile. The algorithm can be set to adapt the beam profile according to several constraints on a target and is applicable to any particle source or beamline. In particular, we showed its application to the 7.5 MeV laser-accelerated proton beam produced at *Laboratoire d'Optique Appliquée* to optimise the transversal dose distribution on a sample. Among different studied configurations we found that a scattering system placed inside the quadrupole transport system flattens the beam profile more efficiently than a scattering system placed after it. This is explained by the fact that such a scattering system flattens the beam profile not only by spreading it but also by taking advantage of the quadrupole chromaticity to intermingle the proton trajectories. We showed that a GA-designed scattering system allows a 20–40% decrease in the standard deviation of the transversal dose distribution at the sample compared to the results obtained by employing a flat foil that uniformly spreads the beam. Such improvement is expected to be even greater for higher energy proton beams, for which the undesired scattering in air and in other materials would be less important. Moreover, our approach can be further extended to improve other parameters such as spectrum or effective LET homogeneity for thicker targets irradiation.

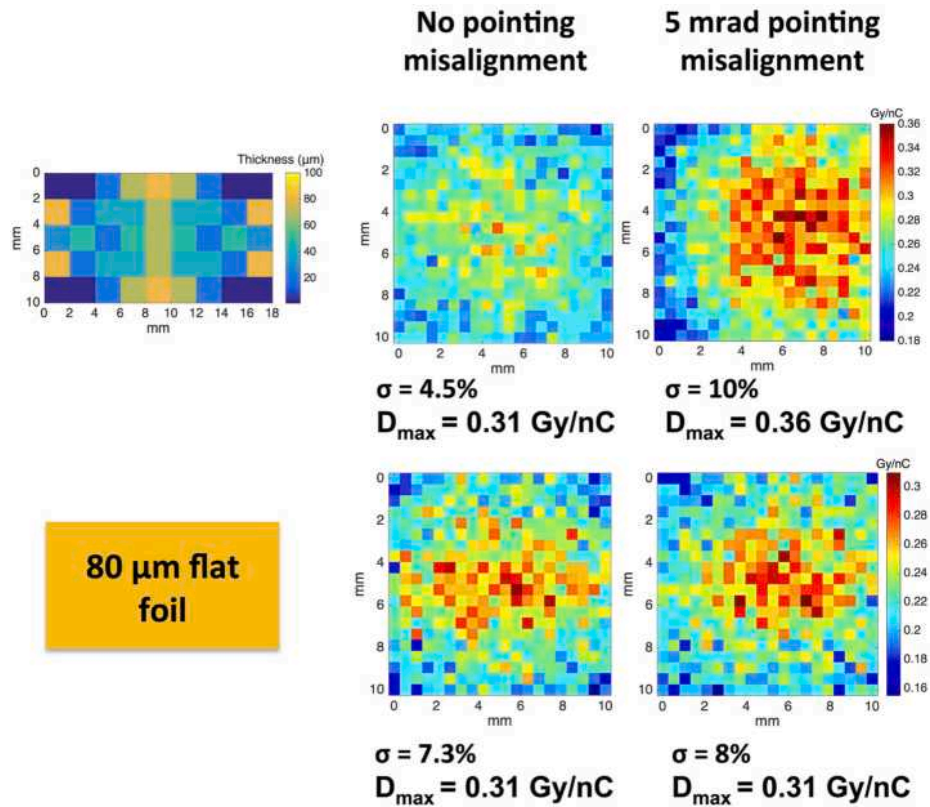

**Fig. 12.** Dose distributions at T2 obtained when employing the GA-designed scattering system placed before the fourth quadrupole and the 80 μm flat scattering system, with no pointing misalignment and with a 5 mrad pointing misalignment.

The scattering systems designed in this study are composed by tiles of discrete thicknesses of Mylar. They can be realized by superposition of 10 μm layers, which are easily handled, or by laser ablation. Also, the realization of such systems to shape higher energy proton beams would be easier thanks to the thicker tiles necessary to introduce the required energy loss and scattering effect. In view of experimental applications, the scattering system sensitivity to typical variations of beam pointing and spectrum was also investigated. The sensitivity of a GA-designed scattering system depends on its position along the beamline, as well as on the spectrum focused at its surface and at the target surface. Therefore, before any experimental application, the instabilities of the beam need to be quantified and a dedicated sensitivity study needs to be carried-out in order to choose the best configuration of the scattering system, allowing not only a high efficiency in ideal conditions but also a low sensitivity to beam instabilities.

**Acknowledgments**

The authors acknowledge the support from the European Union's Horizon 2020 research and innovation program under Grant Agreement No. 654148 Laserlab-Europe and Dr. Emilie Bayart for fruitful and constructive discussions.